# Quantum oscillations and nontrivial topological state in a compensated semimetal TaP$_2$

Hongyuan Wang[1,4,5#], Hao Su[1,4#], Jiuyang Zhang[2], Wei Xia[1,4,5], Yishi Lin[3], Xiaolei Liu[1,4], Xiaofei Hou[1], Zhenhai Yu[1], Na Yu[6], Xia Wang[6], Zhiqiang Zou[6], Yihua Wang[3], Qifeng Liang[7*], Yuhua Zhen[2,*] and Yanfeng Guo[1,4*]

[1]School of Physical Science and Technology, ShanghaiTech University, Shanghai 201210, China

[2]School of Materials Science and Engineering, China University of Petroleum, Qingdao 266580, China

[3]Department of Physics and State Key Laboratory of Surface Physics, Fudan University, Shanghai 200438, China

[4]University of Chinese Academy of Sciences, Beijing 100049, China

[5]Shanghai Institute of Optics and Fine Mechanics, Chinese Academy of Sciences, Shanghai 201800, China

[6]Analytical Instrumentation Center, School of Physical Science and Technology, ShanghaiTech University, Shanghai 201210, China

[7]Department of Physics, University of Shaoxing, Shaoxing 312000, China

We report systematic magneto-transport measurements and *ab initio* calculations on single-crystalline TaP$_2$, a new member of the transition-metal dipnictides. We observed unsaturated magnetoresistance (MR) reaching ~ 700% at a magnetic field (B) of 9 T at 2 K along with striking Shubnikov-de Hass (SdH) oscillations. Our analysis on the SdH oscillations reveals nonzero Berry phase, indicating nontrivial band topology. The analysis also uncovers three fundamental magnetic oscillation frequencies of 72 T, 237 T, and 356 T, consistent with the theoretical calculations which reveal one hole pocket and two electron pockets at the *L* point and one electron pocket at the *Z* point of the Brillouin zone. We also found negative longitudinal MR (*n*-MR) within a narrow window of the angles between B and the electric current (I). The *n*-MR could be fitted with the Adler-Bell-Jackiw chiral anomaly equation but the origin remains yet ambiguous. The *ab inito* calculations suggest TaP$_2$ as a weak topological insulator with the $Z_2$ indices of (0; 111), which exhibits topological surface states on the (001) surface.

PACS numbers:

#These authors contributed equally to this work.
*Corresponding authors:
zhenyh@upc.edu.cn (ZHY),
qfliang@usx.edu.cn(QFL),
guoyf@shanghaitech.edu.cn (YFG).

## I. INTRODUCTION

Nontrivial topological states of matter have been actively pursued in quantum materials due to the associated rich and novel physical properties both in bulk and surface states, such as those found in the intensively studied topological insulators (TIs) [1, 2], Dirac semimetals (DSMs) [3-10], Weyl semimetals (WSMs) [11-20], and topological nodal-line semimetals[21-24], etc. TIs exhibit novel quantum state with metallic edge or surface state within the bulk band gap. The surface state in three-dimensional (3D) TIs is characterized by a linear dispersion of the energy band that could be described by the massless Dirac equation [1, 2]. The unique band structure therefore hosts massless Dirac fermions protected by the time reversal symmetry (TRS). In DSMs and WSMs, two doubly or singly degenerate electronic bands cross each other in the momentum space, thus forming the discrete fourfold Dirac points (DPs) or twofold Weyl points (WPs) and producing exquisite transport properties including high mobility charge carriers [10, 25], large MR [26, 27], chiral anomaly [28-30], anomalous Hall effect [31-33], and optical gyrotropy [34]. The nodal-line semimetals are characterized by the four- or twofold band crossing along 1D lines, rings, links, chains, or knots in momentum space [35-44]. They are expected to display unique physical properties such as flat Landau level (LL) [45], Kondo effect [46], long-range Coulomb interaction [47], and peculiar charge polarization and orbital magnetism [48]. These exotic phenomena have inspired fast growing interest in investigating nontrivial topological states in quantum materials.

Transition-metal dipnictides MPn'$_2$ (M = Ta, and Nb, Pn' = Sb and As) crystallized in the centrosymmetric monoclinic structure (NbAs$_2$-type, space group: $C2/m$) [49] are known for their extremely large MR (XMR) [50-56], where MR is defined as MR = [$\rho$(B) − $\rho$(0)]/$\rho$(0) × 100% in which $\rho$(B) and $\rho$(0) represent the resistivity with and without magnetic field (B), respectively, which even can reach ~ $10^4$% – $10^6$% at B ~ 10 T. The XMR not only offers great potential for use in spintronic devices, but also questions the present theories for MR. To interpret the XMR in MPn'$_2$, it seems that a consensuses relied on the isotropic semi-classical model with perfect electron-hole compensation has been reached [57, 58]. This mechanism can give rise to positive quadratic MR, i. e. MR = $\mu_e\mu_h B^2$ where $\mu_e$ and $\mu_h$ denote the electron and hole carrier mobility, respectively, which seems plausible in interpreting the observed nonsaturating MR. Furthermore, the angle-resolved photoemission spectroscopy (ARPES) measurements on a member of MPn'$_2$, NbAs$_2$, unveiled two electron pockets and two hole pockets with equal volumes in the Brillouin zone (BZ) [56], likely providing compelling evidence in supporting the electron-hole compensation picture. The other parallel interpretation is in connection with the peculiar electronic band topology in topological semimetals (TSMs), which is ascribed to a quantum effect near the linearly dispersed low-energy bulk electrons in topologically protected band structure [59, 60]. Without external B, the backscattering of electrons is strongly suppressed by the peculiar band topology, consequently yielding ultrahigh mobility and long transport lifetime. When B is applied, the lift of the topological protection could give rise to linear field-dependent XMR as is usually observed in some DSMs and WSMs, such as Cd$_3$As$_2$ [10], Na$_3$Bi [30], and MPn [28-30], etc. However, the XMR in some nonmagnetic semimetals such as the nodal-line semimetal CaTX (T = Ag, Cd; X= As, Ge) [61], was suggested to be plausibly explained by evolving both mechanisms simultaneously [26, 62]. In MPn'$_2$, the MR exhibits not only positive parabolic-field-dependent MR but also negative MR (*n*-MR) at low magnetic field and low temperature [54]. The positive MR, for example, in TaAs$_2$ and NbAs$_2$, was experimentally demonstrated to saturate at a sufficient high B not larger than 60 T, likely consistent with the electron-hole compensation picture [53]. However, the saturation behavior needs further examination by measuring to much higher B to confirm the reliability. On top of that, the *n*-MR in MPn'$_2$ could be fitted by the Adler-Bell-Jackiw (ABJ) chiral anomaly equation [54], suggesting that the electron-hole compensation mechanism solely might not account for the complicated behaviors. To achieve a clarification, more experimental evidences are strongly desired. The exploration of new MPn'$_2$ and investigation on the magneto-transport properties are expected to provide valuable clues that help to achieve in-depth insights into the issue. We successfully grew single crystals of a new MPn'$_2$ member, TaP$_2$. We thereafter performed systematic magneto-transport measurements and *ab inito* calculations on high-quality TaP$_2$ single crystals. The results uncover quantum oscillations associated by nontrivial Berry phase, suggesting nontrivial topological nature. Furthermore, angle dependent *n*-MR which could be fitted with the ABJ chiral anomaly equation is also detected within a very narrow angle window at low magnetic field and low temperature.

**II. EXPERIMENTAL**

The TaP$_2$ single crystals were grown by using Te as the flux. Starting materials including tantalum (99.9%), red phosphorus (99.999%) and tellurium (99.999%) were mixed in a molar ratio of Ta : P : Te = 1: 2: 20. The mixture was sealed in a quartz tube, heated up to 1000 ℃ in a furnace and kept at the temperature for 30 hours, then cooled down slowly to 600 ℃ at rate of 2 ℃/h. The assembly was finally taken out of the furnace at 600 ℃ and was put into a centrifuge immediately to remove the excess Bi flux. As we mentioned above, MPn'$_2$ compounds crystallize in the centrosymmetric monoclinic structure with a space group of *C*2/*m*. Assuming the similar structure, the sketch view of the structure of TaP$_2$ should be the one depicted in Fig. 1 (a). To confirm this, by using the typical crystal shown as the inset in Fig. 1(b), we performed careful crystallographic phase and crystal quality examinations on a single-crystal X-ray diffractometer equipped with a Mo Kα radioactive source (λ = 0.71073 Å). The perfect reciprocal space lattice reflections without any other miscellaneous points, seen in Figs. 1(c)-1(e), indicate pure phase and high quality of the crystal used in this study. The diffraction patterns could be satisfyingly indexed on the basis of a monoclinic structure with lattice parameters *a* = 8.8688(16) Å, *b* = 3.2688(5) Å, *c* = 7.4992(14) Å, and *β* = 119.307(7) in the space group *C*2/*m* (No. 12), indicating that our hypothesis is correct. Seen in the structure in Fig. 1(a), the Ta atoms are located inside a trigonal prism formed by the P atoms where the prisms are connected in pairs by sharing one square face. These double prisms are stacked to form infinite columns in the *b* direction. The single crystal XRD characterization confirmed that the largest face of the crystal can be indexed as (001) and the longest side is the [010] direction (the *b*-axis). Element analysis on the randomly selected crystals was performed using energy-dispersive X-ray spectroscopy (EDS). The results confirm good stoichiometry and the free from Bi flux of our crystals (not shown herein). Electrical transport measurements were carried out in a commercial Quantum Design 9 T DynaCool PPMS apparatus. The resistivity was measured in a standard four-probe configuration by attaching Pt wire on the crystals using silver paste. The first principles calculations were performed by using the Vienna *ab initio* simulation package (VASP) [63] and the projected augmented-wave (PAW) potential is adopted [64, 65]. The exchange-correlation functional introduced by Perdew, Burke, and Ernzerhof (PBE) within generalized gradient approximation (GGA) is applied in the calculations [66]. The energy cutoff of the plane-wave basis is set as 520 eV and the forces are relaxed less than 0.01 eV/Å. The positions of atoms were allowed to relax while the lattice constants of the unit cells were fixed to the experimental values.

## III. RESULTS AND DISCUSSION

The temperature (*T*) dependence of longitudinal resistance $R_{xx}$ measured with B perpendicular to the (001) plane and I along the *b*-axis at B = 0 T and 9 T are presented in Fig. 1(b). When B = 0 T, $R_{xx}$ displays an almost linear temperature dependence from 300 K to 75 K and then changes gently upon further cooling down to 2 K, yielding the residual resistivity ratio $R_{xx}$(300 K)/$R_{xx}$(2 K) of approximately 10, indicating an essential semimetallic behavior and high quality of the crystal. With the application of B, $R_{xx}$ is clearly increased and an obvious plateau often found in many topological materials is visible below 100 K. The MR measured at temperatures between 2 and 50 K and at B = 0 – 9 T with B//I//*b* are presented in Fig. 2(a), which display striking quantum oscillations. Moreover, *n*-MR within B ~ 1.3 – 8 T are also observed, which will be discussed later. The pronounced Shubnikov-de Hass (SdH) oscillations at different temperatures derived from the second-order derivatives of $R_{xx}$(T) against the reciprocal magnetic field 1/B are plotted in Fig. 2(b), which are striking at B > 5 T and remain discernible at temperatures at least up to 12 K though the amplitudes systematically decrease with increasing the temperature. The complex periodic behaviors of the amplitudes of the quantum oscillations point to contributions from multiple frequency components. In order to experimentally map out the Fermi surface (FS), the above MR data were further analyzed. The SdH oscillations can be well described by the Lifshitz-Kosevich (LK) formula [67],

$$\Delta\rho_{xx} \propto \frac{2\pi^2 k_B T/\hbar\omega_c}{\sinh(2\pi^2 k_B T/\hbar\omega_c)} \exp(-2\pi^2 k_B T_D/\hbar\omega_c) \cos\left[2\pi\left(\frac{F}{B} - \gamma + \delta\right)\right]$$

where $k_B$ is the Boltzmann constant, $\hbar$ is the Planck's constant, $\omega_c = eB/m^*$ is the cyclotron frequency with $m^*$ denoting the effective cyclotron mass, $T_D$ is the Dingle temperature defined by $T_D = \hbar/2\pi k_B \tau_D$ where $\tau_D$ is the quantum scattering lifetime. The fast Fourier transform (FFT) spectra of the SdH oscillations, shown by the inset in Fig. 2(b), disclose three fundamental frequencies, $F_\alpha$ = 72 T, $F_\beta$ = 237 T, and $F_\gamma$ = 356T, indicating at least three pockets across the Fermi level $E_F$. The experimental result is consistent with the *ab initio* calculations which unveil three types of pockets including one hole pocket and two electron pockets at the *L* point and one electron pockets *Z* point in BZ, respectively. Generally, Hall measurements can provide conceivable information about the carriers. Unfortunately, the Hall measurements on TaP$_2$ were not succeeded limited by the thin stripe shape of the crystals. Further optimization of the growth conditions towards large size crystals is very necessary and the work is on the way. The three fundamental frequencies correspond to the external cross-sectional areas of the FS as $A$ = 0.676 nm$^{-2}$, 2.257 nm$^{-2}$, 3.381 nm$^{-2}$, respectively, which are calculated by using the Onsager relation, $F = (\hbar/2\pi e)A$, where $F$ denotes the frequency of the oscillations. The effective cyclotron mass $m^*$ at $E_F$ can be estimated from the fitting of the temperature dependent FFT amplitudes to the thermal damping term of the LK formula, $\frac{2\pi^2 k_B T/\hbar\omega_c}{\sinh(2\pi^2 k_B T/\hbar\omega_c)}$, as is shown in Fig. 2(c), which gives 0.16 $m_e$, 0.2 $m_e$ and 0.27 $m_e$ corresponding to $F_\alpha$, $F_\beta$, and $F_\gamma$, respectively, where $m_e$ denotes the free electron mass. The corresponding Fermi vectors $k_F$ are 0.464 nm$^{-1}$, 0.848 nm$^{-1}$, 1.038 nm$^{-1}$, respectively, and very large Fermi velocities $v_F$ could consequently be estimated through $v_F = \hbar k_F/m^*$ in the case of an ideal linear dispersion, those are, 3.36×10$^5$ m s$^{-1}$, 4.91×10$^5$ m s$^{-1}$, 4.5×10$^5$ m s$^{-1}$, respectively. The Dingle temperature and the quantum scattering lifetime could be estimated from the inset of Fig. 2(b). Furthermore, the mobility can be also calculated by using the equation $\mu = e\tau_D/m^*$. The results are summarized in Table I.

To gain in-depth insights into the topological nature of the electronic bands, the Landau level (LL) index fan diagram is established aiming to examine the Berry phase of TaP$_2$ accumulated along the cyclotron orbit. This is due to that nontrivial Berry phase is generally considered to be key evidence for Dirac fermions, which is caused by the pseudo-spin rotation under a magnetic field [24, 28]. In the LK equation, γ (= 1/2-φ$_B$/2π) is the Onsager phase factor and δ represents the Fs dimension-dependent correction to the phase shift, which is 0 for two-dimensional (2D) system and ±1/8 for 3D system for nontrivial Berry phase [10, 24]. The LL index phase diagram is shown in Fig. 2(d), in which the ΔR$_{xx}$ valley positions (black box) in 1/B were assigned to be integer indices and the ΔR$_{xx}$ peak positions (red circles) were assigned to be half-integer indices. All the points almost fall on a straight line, thus allowing a linear fitting that gives the intercepts of 0.386(4), 0.43(1), and 0.63(2) corresponding to $F_α$, $F_β$, and $F_γ$, respectively, verifying that all three bands have nonzero Berry phase, as summarized in Table 1. It should be noted that due to that all three frequencies are very large, the nontrivial Berry phase obtained from the LL index fan diagram might not be very accurate. The *ab initio* calculations on TaP$_2$, however, provide complementary evidences for the nontrivial topological nature. Further ARPES measurements are in schedule once the optimization of the crystal size will be successful.

The angle-dependent MR of TaP$_2$ by tilting the direction of B from that perpendicular to the *ab*-plane with I//b at 2 K is presented in Fig. 3. Seen in Fig. 3(a), MR of TaP$_2$ behaves as quadratic at low magnetic field when θ = 0°, indicating the dominate role of orbital contributions and a relatively large MR reaching ~ 700 % at 9 T. When θ is increased, the MR drops quickly and *n*-MR emerges when B is parallel to the current, i.e. θ is 90°. The *n*-MR only exists within a rather narrow θ window, that is, θ = 90°- 93°, as shown in Fig. 3(b). It is clear that the magnitude of *n*-MR at θ = 93° is much stronger than those of other angles, which may due to that the direction of B is only completely parallel with the electrical current at θ = 93° because of the tilting between sample and sample puck. Moreover, the *n*-MR is only visible at low magnetic field, emerging at B ~ 1.3 T and vanishing at B larger than 8 T. The angle dependence of *n*-MR was also observed in other MPn'$_2$ [68], indicating a universal behavior. Previous analysis on the *n*-MR in this family of materials attributed the behavior to the ABJ chiral anomaly. We also repeated the analysis to trace the actual origin for the *n*-MR in TaP$_2$. First, the nonmagnetic nature of TaP$_2$ indicates that the observed *n*-MR has no magnetism source. Second, the *n*-MR display temperature dependence, implying that the current jetting effect associated by inhomogeneous current distribution inside the crystal could also be ruled out. Third, the weak localization effect could not account for the *n*-MR because R$_{xx}$(T) exhibits Fermi-liquid behavior and no any upturn at low temperatures that signifying the effect. Considering the similarities between TaP$_2$ and other MPn'$_2$, the *n*-MR of TaP$_2$ was fitted by using the ABJ chiral anomaly equation which is expressed as $\sigma(B) = (1 + C_w B^2)(\sigma_0 + a\sqrt{B}) + \sigma_N$ [69], where $C_w B^2$ represents the contribution of the chiral current with a positive value, $\sigma_0$ is the zero-field conductivity, $a\sqrt{B}$ is the weak antilocalization contribution with a negative value, and $\sigma_N^{-1} = \rho_0 + A_0 B^2$ denotes the conventional nonlinear band contribution around $E_F$. Fitting result is presented in Fig. 3 (c), showing a nice fit at different temperatures at B = 0 – 3 T. The yielded parameters $\sigma_0$ of 3.45($\pm$0.04) Ω$^{-1}$ and $\rho_0$ of 0.18($\pm$0.03) Ω are consistent with the experimental values. The fitting also gave $C_w$ = 0.14($\pm$0.01)T$^{-2}$, $a$ = -1.35($\pm$0.038) Ω$^{-1}$T$^{-0.5}$, and $A_0$ = 1.2( $\pm$0.2)×10-6 Ω$^{-1}$T$^{-2}$. The value of $A_0$ is very small, indicating that the conventional nonlinear band contribution arisen from the Lorentz force is negligible. This successful fitting to the *n*-MR by using the ABJ anomaly model highlights again the universal behavior in MPn'$_2$, which is tightly related to the nontrivial topological states.

The BZ and calculated band structure of $TaP_2$ are shown in Figs. 4(a) and 4(b), respectively. When spin-orbital coupling (SOC) is neglected, the conduction and valence bands cross at the middle of the high-symmetry path $X_1$ - $Y$ of the BZ and the inclusion of SOC lifts the crossing and opens a gap there. The band structure indicates that the $TaP_2$ is a semimetal, consistent with our resistance data. A hole pocket centered at the $L$ point can be clearly seen, and a tiny electron pocket close to it at the middle of $L$-$I$ path is also observed. The largest carrier pocket is the one located at the $Z$ point where one can find that the bottom of the conduction band has a much lower energy than the one at the $L$ points, which gives to a much large FS. To give a more clear description of these carrier pockets, we also plot the FS of the system in Fig. 5 where the hole pocket at the $L$ points, the two electron pockets around the $L$ points and the largest electron pocket at the $Z$ point are clearly seen. The two electron pockets around $L$ point are symmetrical with each other and the existence of three types of pockets is consistent with our quantum oscillations analysis. The mapped FS in Fig. 5 also confirms that the electron pockets and the hole pockets are almost compensated, thus classifying $TaP_2$ into the compensated semimetal family.

We also calculated the $Z_2$ indices of the crystal and found that the material could be classified as a weak topological insulator (or semimetal) by the indices (0;111). Due to the bulk-edge correspondence, there will be topological edge state at the surfaces of the materials, which could be detected by the ARPES measurement which will be carried out later on optimized crystals. The surface band structure on the (001) surface of $TaP_2$ is illustrated in Fig. 6. The BZ of (001) surface is shown in Fig. 6(a). The configuration of surface density of state in the 2D BZ is plotted in Fig. 6(b) for the Fermi energy. The electron pocket at the $Z$ point is projected onto the point of the 2D BZ and the three pockets near the $L$ point are projected onto the boundary of the BZ. It can be easily found that surface bands connect these Fermi pockets, demonstrating the nontrivial topological electronic structure of the material. The topological surface states are further exhibited in the band structure along the path highlighted in Fig. 6(b) by using dotted line. Furthermore, the band structure of surface electronic state along the dotted line in Fig. 6(b) was also calculated, which is shown in Fig. 6(c).

## IV. CONCLUSION

In summary, we successfully grew single crystals of a new transition-metal dipnictides MPn'$_2$, TaP$_2$. The analysis on the quantum oscillations of the MR indicates the nontrivial band structure. When B is almost parallel to the current, the positive MR shows crossover to *n*-MR which could be described by the ABJ chiral anomaly, strongly highlighting the nontrivial topological states again. Since the calculations also indicate TaP$_2$ as a weak topological insulator, similar as other MPn'$_2$ members, the unusual *n*-MR behavior appeals further investigations to clarify its true origin, for example, by using ARPES to acquire direct information about the band structure. A recent theoretical work on TaAs$_2$ predicted hidden type-II WPs in these centrosymmetric nonmagnetic materials when external magnetic field is applied [70]. The number of WPs depends on the magnitude of Zeeman splitting. It this hypothesis is true, the *n*-MR in MPn'$_2$ could be naturally explained as due to the chirality of the WPs. Though the band structure of MPn'$_2$ is actually very complicated, the discovery of hidden nontrivial topological states would be very valuable in brushing up our knowledge on this family of materials and could pave a way to novel applications of them in devices.

**ACKNOWLEDEMENT**

The authors acknowledge the support by the Natural Science Foundation of Shanghai (Grant No. 17ZR1443300), the Shanghai Pujiang Program (Grant No. 17PJ1406200), and the National Natural Science Foundation of China (Grant No. 11874264, 51772168, 11574215).

**Reference**

1. M. Z. Hasan and C. L. Kane, Rev. Mod. Phys. **82**, 3045 (2010).
2. X.-L. Qi and S.-C. Zhang, Rev. Mod. Phys. **83**, 1057 (2011).
3. Z. Wang, Y. Sun, X.-Q. Chen, C. Franchini, G. Xu, H. Weng, X. Dai, and Z. Fang, Phys. Rev. B **85**, 195320 (2012).
4. Z. Wang, H. M. Weng, Q. Wu, X. Dai, and Z. Fang, Phys. Rev. B **88**, 125427 (2013).
5. Z. Liu, B. Zhou, Y. Zhang, Z. Wang, H. Weng, D. Prabhakaran, S.-K. Mo, Z. Shen, Z. Fang, X. Dai, Z. Hussain, and Y. L. Chen, Science, **343**, 864 (2014).
6. M. Neupane, S. Xu, R. Sankar, N. Alidoust, G. Bian, C. Liu, I. Belopolski, T. Chang, H. Jeng, H. Lin, A. Bansil, F. Chou, and M. Zahid Hasan, Nat. Commun. **5**, 3786 (2014).
7. Z. Liu, J. Jiang, B. Zhou, Z. J. Wang, Y. Zhang, H. M. Weng, D. Prabhakaran, S.-K. Mo, H. Peng, P. Dudin, T. Kim, M. Hoesch, Z. Fang, X. Dai, Z. X. Shen, D. L. Feng, Z. Hussain, and Y. L. Chen, Nat. Mater. **13**, 677 (2014).
8. S. Borisenko, Q. Gibson, D. Evtushinsky, V. Zabolotnyy, B. Büchner, and R. J. Cava, Phys. Rev. Lett. **113**, 027603 (2014).
9. J. Feng, Y. Pang, D. Wu, Z. Wang, H. Weng, J. Li, X. Dai, Z. Fang, Y. Shi, and L. Lu, Phys. Rev. B **92**, 081306(R) (2015).
10. T. Liang, Q. Gibson, M. N. Ali, M. Liu, R. Cava, and N. Ong, Nat. Mater. **14**, 280 (2015).
11. X. G. Wan, A. M. Turner, A. Vishwanath, and S. Y. Savrasov, Phys. Rev. B **83**, 205101 (2011).
12. H. M. Weng, C. Fang, Z. Fang, B. A. Bernevig, and X. Dai, Phys. Rev. X **5**, 011029 (2015).
13. B. Q. Lv, N. Xu, H. M. Weng, J. Z. Ma, P. Richard, X. C. Huang, L. X. Zhao, G. F. Chen, C. E. Matt, F. Bisti, V. N. Strocov, J. Mesot, Z. Fang, X. Dai, T. Qian, M. Shi, and H. Ding, Nat. Phys. **11**, 724 (2015).
14. S. Y. Xu, I. Belopolski, N. Alidoust, M. Neupane, G. Bian, C. L. Zhang, R. Sankar, G. Q. Zheng, Z. J. Yuan, C. C. Lee, S. M. Huang, H. Zheng, J. Ma, D. S. Sanchez, B. K. Wang, A. Bansil, F. C. Chou, P. P. Shibayev, H. Lin, S. Jia, and M. Z. Hasan, Science **349**, 613 (2015).
15. L. X. Yang, Z. K. Liu, Y. Sun, H. Peng, H. F. Yang, T. Zhang, B. Zhou, Y. Zhang, Y. F. Guo, M. Rahn, D. Prabhakaran, Z. Hussain, S.-K. Mo, C. Felser, B. Yan, and Y. L. Chen, Nat. Phys. **11**, 728 (2015).
16. Z. K. Liu, L. X. Yang, Y. Sun, T. Zhang, H. Peng, H. F. Yang, C. Chen. Y. Zhang, Y. F. Guo, D. Prabhakaran, M. Schmidt, Z. Hussain, S.-K. Mo, C. Felser, B. Yan, and Y. L. Chen, Nat. Mater. **15**, 27 (2016).
17. S.-Y. Xu, N. Alidoust, I. Belopolski, Z. Yuan, G. Bian, T.-R. Chang, H. Zheng, V. N. Strocov, D. S. Sanchez, G. Chang, C. Zhang, D. Mou, Y. Wu, L. Huang, C.-C. Lee, S.-M. Huang, B. Wang, A. Bansil, H.-T. Jeng, T. Neupert, A. Kaminski, H. Lin, S. Jia, and M. Zahid Hasan, Nat. Phys. **11**, 748 (2015).
18. Y. Sun, S.-C. Wu, M. N. Ali, C. Felser, and B. Yan, Phys. Rev. B **92**, 161107(R)

(2015).

19. L. Huang, T. M. McCormick, M. Ochi, Z. Zhao, M. T. Suzuki, R. Arita, Y. Wu, D. Mou, H. Cao, J. Yan J, N. Trivedi, and A. Kaminski, Nat. Mater. **15**, 1155 (2016).
20. K. Deng, G. L. Wan, P. Deng, K. N. Zhang, S. J. Ding, E.Y. Wang, M. Z. Yan, H. Q. Huang, H. Y. Zhang, Z. L. Xu, J. Denlinger, A. Fedorov, H. T. Yang, W. H. Duan, H. Yao, Y. Wu, S. S. Fan, H. J. Zhang, X. Chen, and S. Y. Zhou, Nat. Phys. **12**, 1105 (2016).
21. Y. Wu, L. L. Wang, E. Mun, D. D. Johnson, D. X. Mou, L. N. Huang, Y. B. Lee, S. L. Bud'ko, P. C. Canfeild, and A. Kaminski, Nat. Phys. **12**, 667 (2016).
22. G. Bian, T.-R. Chang, R. Sankar, S.-Y. Xu, H. Zheng, T. Neupert, C.-K. Chiu, S.-M. Huang, G. Chang, I. Belopolski, D. S. Sanchez, M. Neupane, N. Alidoust, C. Liu, B. Wang, H.-T. Jeng, A. Bansil, F. Chou, H. Lin, and M. Z. Hasan, Nat. Commun. **7**, 10556 (2016).
23. L. M. Schoop, M. N. Ali, C. Straßer, A. Topp, A. Varyhalov, D. Marchenko, V. Duppel, S. S. S. Parkin, B. V. Lotsch, and C. R. Ast, and Nat. Commun. **7**, 11696 (2016)
24. J. Hu, Z. J. Tang, J. Y. Liu, X. Liu, Y. L. Zhu, D. Graf, K. Myhro, S. Tran, C. N. Lau, J. Wei, and Z. Q. Mao, Phys. Rev. Lett. **117**, 016602 (2016).
25. C. Shekhar, A. K. Nayak, Y. Sun, M. Schmidt, M. Nicklas, I. Leermakers, U. Zeitler, Z. K. Liu, Y. L. Chen, W. Schnelle, J. Grin, C. Felser, and B. Yan, Nat. Phys. **11**, 645 (2015).
26. M. N. Ali, J. Xiong, S. Flynn, J. Tao, Q. D. Gibson, L. M. Schoop, T. Liang, N. Haldolaarachchige, M. Hirschberger, N. P. Ong, and R. J. Cava, Nature **514**, 205 (2014).
27. Z. Zhu, X. Lin, J. Liu, B. Fauqué, Q. Tao, C. Yang, Y. Shi, and K. Behnia, Phys. Rev. Lett. **114**, 176601 (2015).
28. C.-L. Zhang, S.-Y. Xu, I. Belopolski, Z. Yuan, Z. Lin, B. Tong, G. Bian, N. Alidoust, C.-C. Lee, S.-M. Huang, T.-R. Chang, G. Chang, C.-H. Hsu, H.-T. Jeng, M. Neupane, D. S. Sanchez, H. Zheng, J. Wang, H. Lin, C. Zhang, H.-Z. Lu, S.-Q. Shen, T. Neupert, M. Z. Hasan, and S. Jia, Nat. Commun. **7**, 10735 (2016).
29. X. C. Huang, L. X. Zhao, Y. J. Long, P. P. Wang, D. Chen, Z. H. Yang, H. Liang, M. Q. Xue, H. Weng, Z. Fang, X. Dai, and G. F. Chen, Phys. Rev. X **5**, 031023 (2015).
30. J. Xiong, S. K. Kushwaha, T. Liang, J. W. Krizan, M. Hirschberger, W. Wang, R. J. Cava, and N. P. Ong, Science **350**, 413 (2015).
31. K.-Y. Yang, Y.-M. Lu, and Y. Ran, Phys. Rev B **84**, 075129 (2011).
32. E. K. Liu, Y. Sun, N. Kumar, L. Muechler, A. L. Sun, L. Jiao, S. Y. Yang, D. F. Liu, A. J. Liang, Q. N. Xu, J. Kroder, V. Süß, H. Borrmann, C. Shekhar, Z. S. Wang, C. Y. Xi, W. H. Wang, W. Schnelle, S. Wirth, Sebastian T. B. Goennenwein, and C. Felser, Nat. Phys.**14**, 1125 (2018).
33. Q. Wang, Y. F. Xu, R. Lou, Z. H. Liu, M. Li, Y. B. Huang, D. W. Shen, H. M. Weng, S. C. Wang, and H. C. Lei, Nat. Commun. **9**, 3681 (2018).
34. S. Zhong, J. Orenstein, and J. E. Moore, Phys. Rev. Lett. **115**, 117403 (2015).
35. T. Bzdusek, Q. Wu, A. Riiegg, M. Sigrist, and A. A. Soluyanov, Nature (London)

**538**, 75 (2016).

36. X. Feng, C. M. Yue, Z. D. Song, Q. S. Wu, and B. Wen, Phys. Rev. Mater. **2**, 014202 (2018).
37. Q.-F. Liang, J. Zhou, R. Yu, Z. Wang, and H.-M. Weng, Phys. Rev B **93**, 085427 (2016).
38. J. Lian, L. Yu, Q.-F. Liang, J. Zhou, R. Yu, and H.-M. Weng, npj Comput. Mater. **5**, 10 (2019).
39. X. M. Zhang, Z.-M. Yu, X.-L. Sheng, H. Y. Yang, and S. A. Yang, Phys. Rev B **95**, 235116 (2017).
40. W. Chen, H.-Z. Lu, and J.-M. Hou, Phys. Rev B **96**, 041102(R) (2017).
41. Z. Yan, R. Bi, H. Shen, L. Lu, S.-C. Zhang, and Z. Wang, Phys. Rev B **96**, 041103(R) (2017).
42. M. Ezawa, Phys. Rev. B **96**, 041202(R) (2017).
43. P.-Y. Chang and C.-H. Yee, Phys. Rev. B **96**, 081114(R) (2017).
44. Y. Zhou, F. Xiong, X. G. Wan, and J. An, Phys. Rev. B **97**, 155140 (2018).
45. J.-W. Rhim and Y. B. Kim, Phys. Rev B **92**, 045126 (2015).
46. A. K. Mitchell and L. Fritz, Phys. Rev. B **92**, 121109(R) (2015).
47. Y. Huh, E.G. Moon, and Y. B. Kim, Phys. Rev B **93**, 035138 (2016).
48. S. T. Ramamurthy and T. L. Hughes, Phys. Rev. B **95**, 075138 (2017).
49. W. Jeitschko and P. C. Donohue, Acta Cryst. **B29**, 783 (1973).
50. D. S. Wu, J. Liao, W. Yi, X. Wang, P. G. Li, H. M. Weng, Y. g. Shi, Y. Q. Li, J. L. Luo, X. Dai, and Z. Fang, Appl. Phys. Lett. **108**, 042105 (2016).
51. K. Wang, D. Graft, L. Li, L. Wang, and C. Petrovic, Sci. Rep. **4**, 7328 (2014).
52. Z. Yuan, H. Lu, Y. Liu, J. Wang, and S. Jia, Phys. Rev. B **93**, 184405 (2016).
53. C.-L. Zhang, Z. Yuan, Q.-D. Jiang, B. Tong, C. Zhang, X. C. Xie, and S. Jia, Phys. Rev. B **95**, 085202 (2017).
54. Yongkang Luo, R. D. McDonald, P. F. S. Rosa, B. Scott, N. Wakeham, N. J. Ghimire, E. D. Bauer, J. D. Thompson, and F. Ronning, Sci. Rep. 6, 27294 (2016).
55. Y.-Y. Wang, Q.-H. Yu, P.-J. Guo, K. Liu, and T.-L. Xia, Phys. Rev. B **94**, 041103(R) (2016).
56. B. Shen, X. Deng, G. Kotliar, and N. Ni, Phys. Rev. B **93**, 195119 (2016).
57. J. F. He, C. F. Zhang, N. J. Ghimire, T. Liang, C. J. Jia, J. Jiang, S. J. Tang, S. Chen, Y. He, S.-K. Mo, C. C. Hwang, M. Hashimoto, D. H. Lu, B. Moritz, T. P. Devereaux, Y. L. Chen, J. F. Mitchell, and Z.-X. Shen, Phys. Rev. Lett. **117**, 267201 (2016).
58. L.-K. Zeng, R. Lou, D.-S. Wu, Q. N. Xu, P.-J. Guo, L.-Y. Kong, Y.-G. Zhong, J.-Z. Ma, B.-B. Fu, P. Richard, P. Wang, G. T. Liu, L. Lu, Y. B. Huang, C. Fang, S.-S. Sun, Q. Wang, L. Wang, Y.-G. Shi, H. M. Weng, H.-C. Lei, K. Liu, S.-C. Wang, T. Qian, J.-L. Luo, and H. Ding, Phys. Rev. Lett. **117**, 127204 (2016).
59. A. A. Abrikosov, Phys. Rev. B **58**, 2788 (1998).
60. A. A. Abrikosov, Europhys. Lett. **49**, 789 (2000).
61. E. Emmanouilidou, B. Shen, X. Deng, T.-R. Chang, A. Shi, G. Kotliar, S.-Y. Xu, and N. Ni, Phys. Rev. B **95**, 245113 (2017).
62. X. Du, S. W. Tsai, D. L. Maslov, and A. F. Hebard, Phys. Rev. Lett. **94**, 166601

(2005).
63. G. Kresse and J. Furthmüller, Phys. Rev. B **54**, 11169 (1996).
64. P. E. Blöchl, Phys. Rev. B **50**, 17953 (1994).
65. G. Kresse and D. Joubert, Phys. Rev. B **59**, 1758 (1999).
66. J. P. Perdew, K. Burke, and M. Ernzerhof, Phys. Rev. Lett. **77**, 3865 (1996).
67. E. M. Lifshits and A. M. Kosevich, J. Phys. Chem. Solids **4**, 1(1958).
68. C. M. Wang, H. Z. Lu, and S. Q. Shen, Phys. Rev. Lett. **117**, 077201 (2016).
69. H. J. Kim, K. S. Kim, J. F. Wang, M. Sasaki, N. Satoh, A. Ohnishi, M. Kitaura, M. Yang, and L. Li, Phys. Rev. Lett. **111**, 246603 (2013).
70. D. Gresch, Q. S. Wu, G. W. Winkler, A. A. Soluyanov. New. J. Phys. **19**, 035001 (2017).

TABLE I

| F (T) | $A$ (nm$^{-2}$) | $k_F$ (nm$^{-1}$) | $v_F$ (m/s) | $E_F$ (mev) | $m^*/m_e$ | $T_D$ (K) | $\tau_D$ (s) | $\mu$ (cm$^2$/Vs) | Berry phase |
|---|---|---|---|---|---|---|---|---|---|
| 72 | 0.686 | 0.467 | $3.39 \times 10^5$ | 104.35 | 0.16 | 6 | $2 \times 10^{-13}$ | 2235 | 0.76 π |
| 237 | 2.257 | 0.848 | $4.91 \times 10^5$ | 274.28 | 0.20 | 14.53 | $8.37 \times 10^{-14}$ | 736.24 | 0.86 π |
| 356 | 3.390 | 1.039 | $4.46 \times 10^5$ | 305.75 | 0.27 | 25.10 | $4.85 \times 10^{-14}$ | 315.88 | 1.26 π |

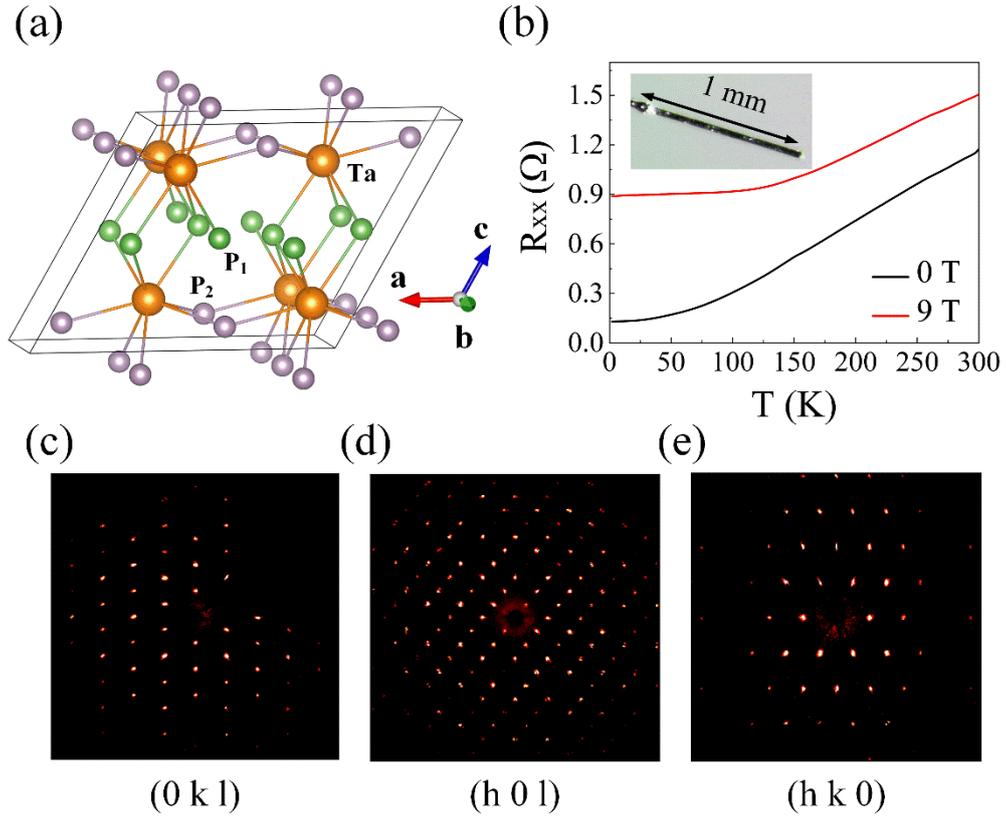

FIG. 1. (a) The schematic crystal structure of TaP$_2$. (b) The longitudinal resistance measured between 2 – 300 K at B = 0 and 9 T. Inset shows the optical image of the typical TaP$_2$ single crystal used in this work. (c)-(e) The diffraction patterns in the reciprocal space along (0kl), (h0l), and (hk0) directions.

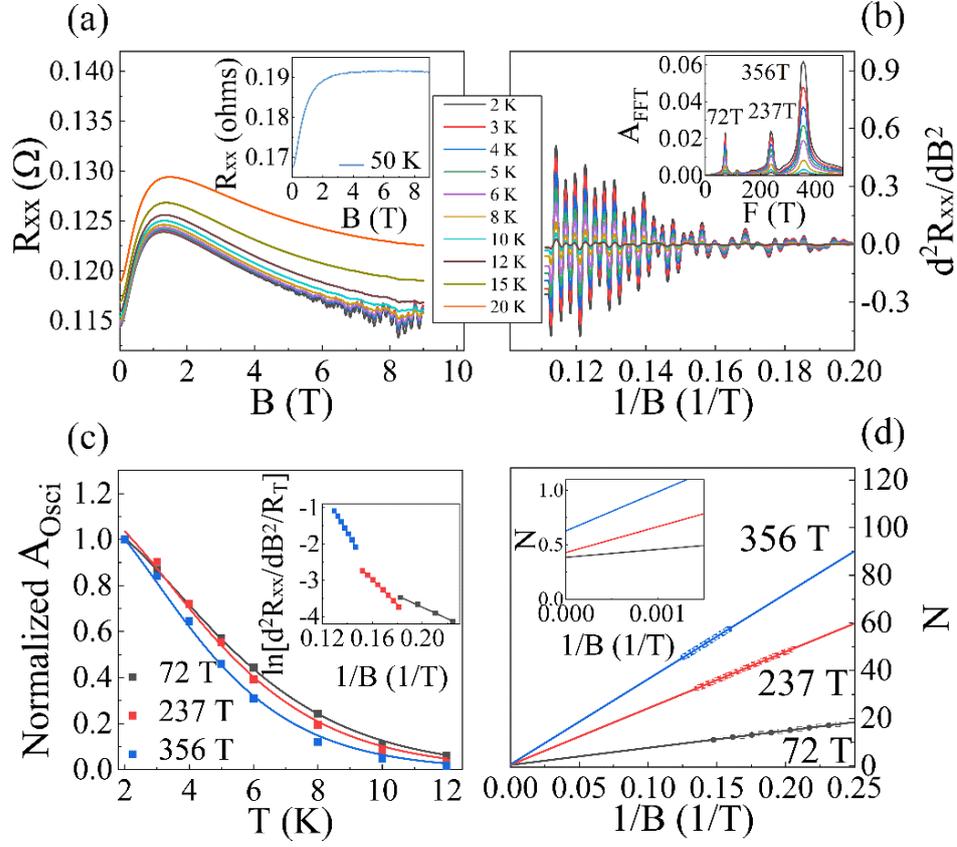

FIG. 2. (a) The longitudinal MR $R_{xx}$ versus magnetic field B between $T = 2 - 50$ K. Inset shows the disappearance of *n*-MR at 50 K. (b) The SdH oscillatory component as a function of 1/B obtained from the second-order derivatives. The inset shows the obtained three main frequencies of 71 T, 238 T and 355 T. (c) The temperature dependence of relative amplitudes of the SdH oscillations for the three frequencies. The solid lines denote the fitting by using the L-K formula. (d) The Landau-level indices extracted from the SdH oscillations plotted as a function of 1/B. The solid lines indicate the linear plots to the data.

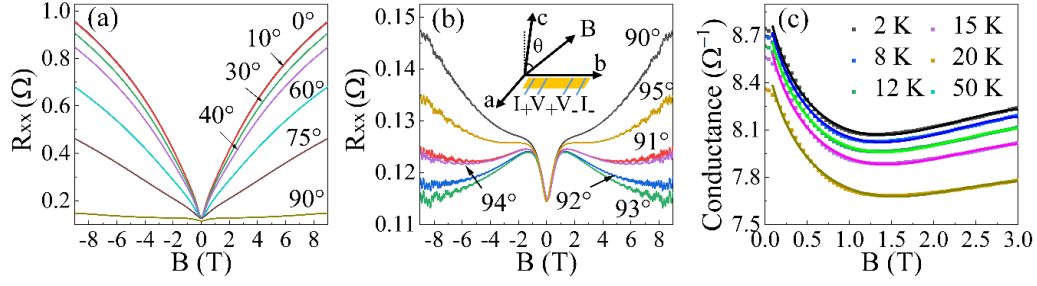

FIG. 3. (a) The longitudinal MR $R_{xx}$ versus B at 2 K and different angles θ varying from 0° to 90°. (b) The $R_{xx}$ versus B at 2 K and θ = 90° to 95°. Inset shows the schematic measurement configuration. The SdH oscillatory component as a function of 1/B obtained after subtracting a smooth background at different angle. (c) Results of the fitting to conductance at different temperatures between 0 T – 3 T by the chiral anomaly equation (solid lines).

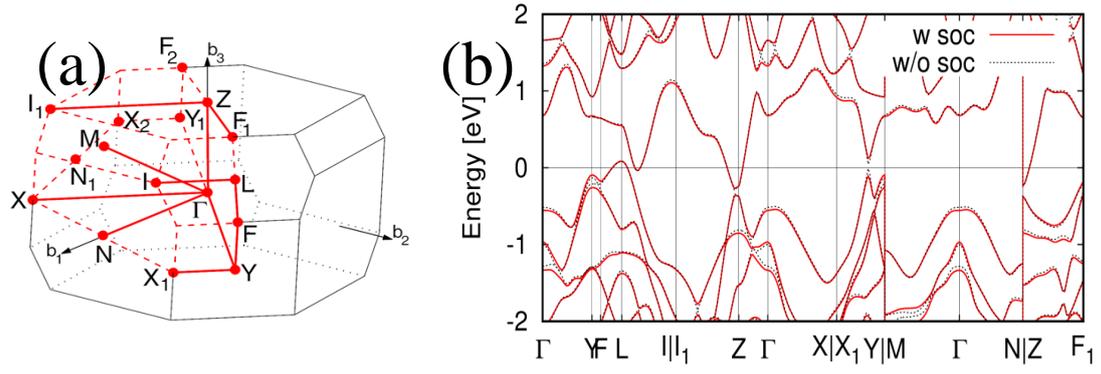

FIG. 4. (a) Brilloiun zone of TaP$_2$ in the monoclinic structure. (b) Band structures of TaP$_2$. The solid (red) and dotted (black) curves represent the band structures of TaP$_2$ with and without spin-orbital coupling, respectively.

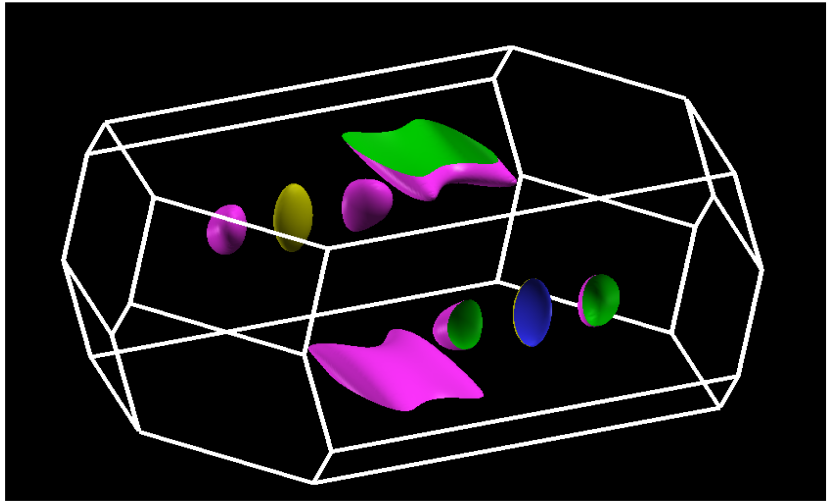

FIG. 5. Fermi surface of TaP$_2$.

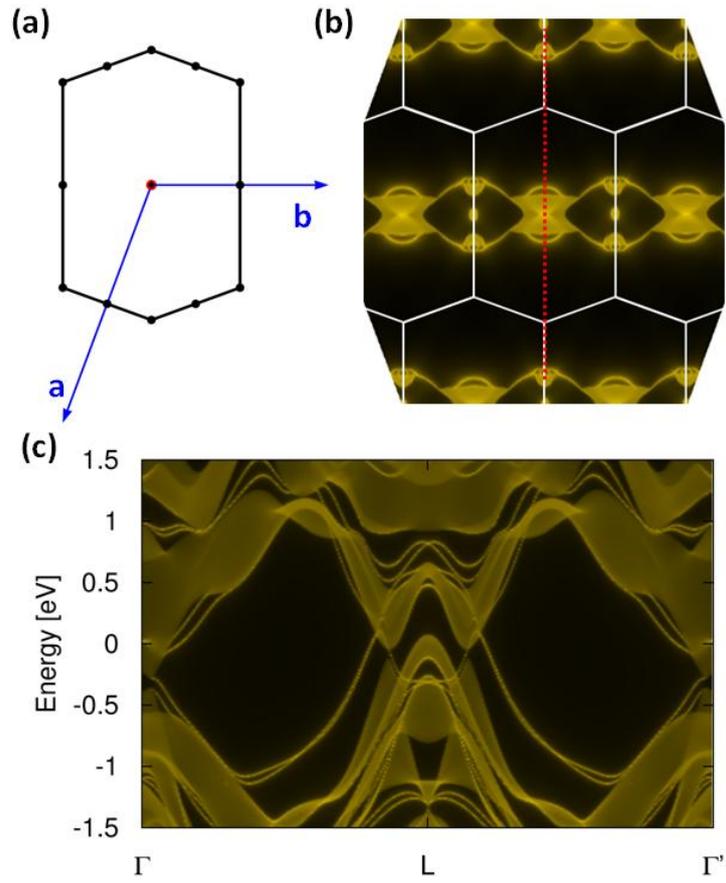

FIG. 6. Surface electronic structure of TaP$_2$ on (001) surface. (a) Brillouin zone of (001) surface. (b) Surface density of states. (c) Band structure of surface electronic state along the dotted line in (b).